%% file: main.tex
\pdfoutput=1
\documentclass{article}
\usepackage{cite}
\usepackage{spconf,amsmath,graphicx,amsfonts}
\usepackage{algorithm}
\usepackage{algorithmic}
\usepackage{hyperref}
\usepackage[numbers,sort&compress]{natbib}
\usepackage{multirow}
\usepackage{comment}
\usepackage{booktabs}
\usepackage{subfigure}
\usepackage{arabtex}

\title{Adversarial sample detection for speaker verification \\ by neural vocoders}

\makeatletter
\def\name#1{\gdef\@name{#1\\}}
\makeatother
\name{\em{Haibin Wu$^{1}$,
    Po-chun Hsu$^{1}$,
    Ji Gao$^{2}$,
    Shanshan Zhang$^{2}$,
    Shen Huang$^{2}$,
    Jian Kang$^{2}$,} \\
    \em{Zhiyong Wu$^{3}$,
    Helen Meng$^4$,
    Hung-yi Lee$^{1}$}
}

\address{
  $^1$ Graduate Institute of Communication Engineering, National Taiwan University \\
  $^4$ Centre for Perceptual and Interactive Intelligence, The Chinese University of Hong Kong \\
  $^3$ Shenzhen International Graduate School, Tsinghua University \\
  $^2$ Tencent Research, Beijing, China 
}
 
\begin{document}
\ninept

\maketitle
%
\input{0-abstract-keywords}
\input{1-introduction}
\input{2-method}

\input{3-experimental-setup}

\input{4-experimental-results}

\input{5-conclusion}


\input{main.bbl}
\bibliographystyle{IEEEbib}

\end{document}

%% file: 0-abstract-keywords.tex
\begin{abstract}
Automatic speaker verification (ASV), one of the most important technology for biometric identification, has been widely adopted in security-critical applications.
However, ASV is seriously vulnerable to recently emerged adversarial attacks, yet effective countermeasures against them are limited.
In this paper, we adopt neural vocoders to spot adversarial samples for ASV.
We use the neural vocoder to re-synthesize audio and find that the difference between the ASV scores for the original and re-synthesized audio is a good indicator for discrimination between genuine and adversarial samples.
This effort is, to the best of our knowledge, among the first to pursue such a technical direction for detecting time-domain adversarial samples for ASV, and hence there is a lack of established baselines for comparison.
Consequently, we implement the Griffin-Lim algorithm as the detection baseline. 
The proposed approach achieves effective detection performance that outperforms the baselines in all the settings. 
We also show that the neural vocoder adopted in the detection framework is dataset-independent.
Our codes will be made open-source for future works to do fair comparison \footnote[1]{https://github.com/HaibinWu666/spot-adv-by-vocoder. This work was done while Haibin Wu was an intern at Tencent Research, Beijing.}.
\end{abstract}

%% file: 1-introduction.tex
\section{Introduction}
\label{sec:intro}
ASV refers to verifying whether a utterance is uttered by a certain person, and has been adopted in a wide range of security-critical applications
Recently, deep learning has dramatically boosted advancements in ASV, resulting in a variety of high-performance ASV models \cite{dehak2010front,kenny2012small,snyder2018x}.
However, ASV is susceptible to the recently emerged adversarial attacks \cite{kreuk2018fooling,das2020attacker}, causing serious security problems.

Using adversarial samples to attack machine learning models is called adversarial attack \cite{szegedy2013intriguing}.
Adversarial samples are similar to their genuine counterparts according to human perception, but can can fool high-performance models, which is surprising.
Speech processing models, including automatic speech recognition (ASR) \cite{carlini2018audio,qin2019imperceptible} and anti-spoofing for ASV \cite{liu2019adversarial,wu2020defense_2,wu2020defense}, are also susceptible to adversarial attacks.
ASV models are no exception \cite{das2020attacker,jati2021adversarial} – the first illustration of the vulnerability of ASV models to adversarial attacks was presented in \cite{kreuk2018fooling}. 
Also, state-of-the-art ASV models, including i-vector and x-vector models, can be manipulated by adversarial samples \cite{villalba2020x,li2020adversarial}.
This is followed by efforts that investigate more malicious adversarial attacks from the perspectives of universality \cite{marras2019adversarial}, in-the-air transferability \cite{li2020practical}, and imperceptibility \cite{wang2020inaudible}.

However, due to limited research efforts \cite{wang2019adversarial,li2020investigating,zhang2020adversarial,wu2021adversarialasv,wu2021improving,joshi2021adversarial,wu2021voting} in adversarial defense on ASV, effective strategies remain an open question. 
Wang et al. \cite{wang2019adversarial} adopts adversarial training to mitigate adversarial attacks for ASV by injecting adversarial data into the training set.
Li et al. \cite{li2020investigating} proposes a detection model for adversarial samples by training it on a mixture of adversarial samples and genuine samples.
Zhang et al. \cite{zhang2020adversarial} harnesses an independent DNN filter trained with adversarial samples and applies it to purify the adversarial samples.
However, the above methods \cite{wang2019adversarial,li2020investigating,zhang2020adversarial} require the knowledge of the attack algorithms used by attackers.
It is impractical to assume that the ASV system designers know in advance which attack algorithms will be implemented by attackers in-the-wild, not to mention that such methods \cite{wang2019adversarial,li2020investigating,zhang2020adversarial} may overfit to a specific adversarial attack algorithm.
Self-supervised learning models have also been proposed \cite{wu2021adversarialasv,wu2021improving} as a filter to purify the adversarial noise.
Josshi et al. \cite{joshi2021adversarial} proposes four pre-processing defenses, and \cite{wu2021voting} introduces the idea of voting to prevent risky decisions of ASV when encountering adversarial samples.

We propose neural vocoders to detect adversarial samples and use Parallel WaveGAN \cite{yamamoto2020parallel} as a case study. 
Vocoders are usually adopted to attack ASV systems by generating spoofing audios \cite{todisco2019asvspoof}.
We propose the contrary to harness the vocoders to defend ASV systems. 
Defense aims at purifying the adversarial noise, while detection aims at spotting adversarial samples and filtering them away.
In contrast to the approaches which need to know the adversarial attack methods \cite{wang2019adversarial,li2020investigating,zhang2020adversarial}, the proposed approach does not require such information.
Compared to approaches that defend ASV in the frequency domain \cite{wu2021adversarialasv,wu2021improving}, the proposed approach directly detects adversarial samples in the time domain.
So the proposed method can serve as a complement of \cite{wu2021adversarialasv,wu2021improving}.
The vocoder in \cite{joshi2021adversarial} is for defense, while the proposed approach is for detection.
Wu et al. \cite{wu2021voting} focuses on defense, yet this paper aims at detection. 
To the best of our knowledge, this is the first paper to adopt neural vocoders to detect time-domain adversarial samples for ASV, and our results demonstrate effectiveness over the traditional Griffin-Lim vocoder.

%% file: 2-method.tex
\section{Background}
\subsection{Automatic speaker verification}
The objective of ASV is to authenticate the claimed identity of a speaker by a piece of his/her speech and some enrolled speaker records.
The procedure of ASV can be divided into feature engineering, speaker embedding extraction, and similarity scoring. 
Feature engineering aims at transforming a piece of utterance in waveform representation, into acoustic features, such as Mel-frequency cepstral coefficients (MFCCs), filter-banks, and spectrograms. 
The speaker embedding extraction procedure of recently ASV models \cite{dehak2010front,kenny2012small,snyder2018x} usually extracts utterance-level speaker embedding from acoustic features.
Then similarity scoring will measure the similarity between the testing speaker embedding and the enrolled speaker embedding.
The higher the score, the more likely that the enrolment utterance and the testing utterance belong to the same speaker, and vice versa. 
Let us denote the testing utterance and the enroll utterance as $x_{t}$ and $x_{e}$ respectively.
For simplicity, we combine the above three procedures and view ASV as an end-to-end function $f$: 
\begin{align}
    &s = f(x_{t}, x_{e}), 
\end{align}
where $s$ is the similarity score between $x_{t}$ and $x_{e}$.

\subsection{Adversarial attack}
Attackers deliberately incorporate a tiny perturbation, which is indistinguishable from human perception, and combine it with the original sample to generate the new sample, which will manipulate the model give wrong prediction.
The new sample and the tiny perturbation are denoted as the adversarial sample and adversarial noise, respectively.
Suppose that the attackers in the wild have access to the internals of the ASV system, including structures, parameters and gradients, and have the access to the testing utterance $x_{t}$.
They aim at crafting such an adversarial utterance by finding an adversarial perturbation.
Different searching strategies for elaborating adversarial noise result in different attack algorithms.
In this work, we adopt a powerful attack method, the basic iterative method (BIM) \cite{kurakin2016adversarial}.
During BIM attack, attackers will start from $x_{t}^{0}=x_{t}$, then iteratively update it to find the adversarial sample:
\begin{equation}
\begin{aligned}
    x_{t}^{k+1}=clip\left(x_{t}^{k} + \alpha \cdot (-1)^{is\_tgt} \cdot sign\left(\nabla_{x_{t}^{k}}f(x_{t}^{k}, x_{e}) \right)\right), 
    \\ for \, k=0,1, \ldots, K-1,
\end{aligned}
\end{equation}
where $clip(.)$ is the clipping function which make sure that $||x_{t}^{k+1} - x_{t}||_{\infty}\leq \epsilon$,
$\epsilon$, denotes the attack budget or intensity predefined by the attackers,
$\epsilon \geq 0 \in \mathbb{R}$,
$\alpha$ is the step size,
$is\_tgt=1$ and $is\_tgt=0$ for the target trial and the non-target trial respectively,
$K$ is the number of iterations and we define $K = \lceil \epsilon / \alpha \rceil$, where $\lceil.\rceil$ denotes the ceiling function.
In target trials, the testing and enrolment utterances are pronounced by the same speaker.  In non-target trials, they belong to different speakers.
Take the non-target trial as an example -- after the BIM attack, the similarity score between the testing and enrolment utterances will be high, which will mislead the ASV system to falsely accept the imposter.
We recommend that our readers listen to the demo of the deliberately crafted adversarial samples \footnote[2]{https://haibinwu666.github.io/adv-audio-demo/index.html}, which tend to be indistinguishable from their genuine counterparts.

\subsection{Vocoder}
Due to the lack of phase information, speech waveforms cannot be restored directly from acoustic features, such as linear spectrograms and mel-spectrograms.
The traditional vocoder, Griffin-Lim, \cite{griffin1984signal} is usually used to reconstruct phase information.
However, it inevitably introduces distortion during reconstruction, resulting in reduced speech quality.
We argue that the introduced distortion may also degrade the effect of the attack.
Another approach, the neural vocoder, takes acoustic features as conditions and uses a neural network to generate speech signals.
Since a neural vocoder is trained to maximize the likelihood of real speech in training data, we expect that when given distorted or attacked acoustic features, the neural vocoder can generate their genuine counterparts. 
In contrast to Griffin-Lim, a neural vocoder is a data-driven method, which can model the manifolds of genuine data, and thus generates waveform with lowered distortion. 

Neural vocoders can restore high-quality speech but with slow inference speed due to the autoregressive architecture.
Parallel WaveGAN \cite{yamamoto2020parallel} adopted a model based on dilated CNN, which can generate audio samples in parallel.
They jointly trained the model using the adversarial loss in GAN and the proposed loss on the frequency domain.
Parallel synthesis improves the efficiency of speech generation, while the GAN architecture can make the Parallel WaveGAN effectively model the distribution of real speech.
Thus in this work, we adopt Parallel WaveGAN for spotting adversarial samples.

\section{Neural vocoder is all you need}
\label{sec:method}
\begin{figure}[ht]
  \centering
  \centerline{\includegraphics[width=\linewidth]{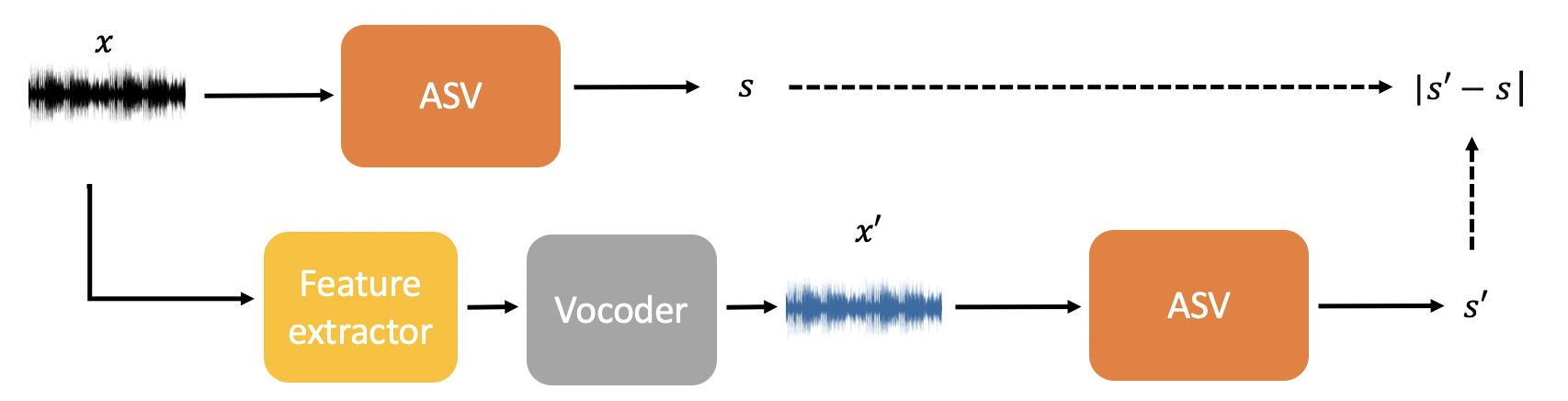}}
  \vspace{-5pt}
  \caption{Proposed detection framework. $s$ and $s'$ are the ASV scores for $x$ and $x'$. $|s-s'|$ is the absolute value between $s$ and $s'$.}
  \label{fig:method}
  \vspace{-5pt}
\end{figure}

\subsection{The detection procedure}
\label{subsec:vocoder based detection}
We first detail the detection procedure, followed by the reason why it works.
The vocoder\footnote[3]{Unless specified otherwise, the use of “vocoder” refers to the “neural vocoder” in the following sections.}-based detection framework is shown in Fig.~\ref{fig:method}.
For brevity, we omit the enrollment utterance $x_{e}$.
The subscript of $x_{t}$ is also omitted, and we use $x$ to denote the testing utterance.
We use $x'$ to denote the testing utterance after feature extraction and vocoder preprocessing (yellow block and gray block in Fig.~\ref{fig:method}).
We follow the procedure in Fig.~\ref{fig:method}, and get $|s-s'|$ for a piece of testing utterance $x$.
Denote the score variation $d=|s-s'|$.
Denote $\mathbb{T}_{gen} = \{x_{gen}^{1}, x_{gen}^{2},...,x_{gen}^{I}\}$ is the set of genuine testing utterances, and $\vert \mathbb{T}_{gen} \vert $ denotes the number of elements in set $\mathbb{T}_{gen}$.
Then we derive $\{d_{gen}^{1}=|s_{gen}^{1}-{s_{gen}^{1}}'|,d_{gen}^{2}=|s_{gen}^{2}-{s_{gen}^{2}}'|, ...,d_{gen}^{I}=|s_{gen}^{I}-{s_{gen}^{I}}'|\}$ for $\mathbb{T}_{gen}$ as shown in Fig.~\ref{fig:method}, where $s_{gen}^{i}$ and ${s_{gen}^{i}}'$ are the ASV scores for $x_{gen}^{i}$ before and after vocoder preprocessing respectively.
Given a false positive rate for detection ($FPR_{given}$, a real number), such that $FPR_{given} \in [0,1]$, for genuine samples, we derive a detection threshold $\tau_{det}$:
\begin{align}
    &FPR_{det}(\tau) = \frac{\vert \{ d_{gen}^{i} > \tau : x_{gen}^{i} \in \mathbb{T}_{gen} \} \vert}{\vert \mathbb{T}_{gen} \vert} \label{eq:det-far} \\
    &\tau_{det} = \{ \tau \in \mathbb{R} : FPR_{det}(\tau) =FPR_{given} \} \label{eq:det-threshold} 
\end{align}
where $FPR_{det}(\tau)$ is the false positive rate for genuine samples given a threshold $\tau$, $d_{gen}^{i}$ is derived by $x_{gen}^{i}$ as shown in Fig.~\ref{fig:method}.
In realistic conditions, the ASV system designer is unaware of adversarial samples, not to mention which exact adversarial attack algorithm will be adopted.
So the detection threshold $\tau_{det}$ is determined based on genuine samples.
Hence the detection method does not require knowledge of adversarial sample generation.

Given a testing utterance, be it adversarial or genuine, $|s-s'|$ will be derived, and the system will label it adversarial if $|s-s'| > \tau_{det}$, and vice versa.
The detection rate ($DR_{\tau_{det}}$) under $\tau_{det}$, which is determined by Eq.~\ref{eq:det-threshold}, for adversarial data can be derived as:
\begin{align}
    &DR_{\tau_{det}} = \frac{\vert \{ d_{adv}^{i} > \tau_{det} : x_{adv}^{i} \in \mathbb{T}_{adv} \} \vert}{\vert \mathbb{T}_{adv} \vert} \label{eq:det-rate} 
\end{align}
where $\mathbb{T}_{adv}$ denotes the set of adversarial testing utterances, and $d_{adv}^{i}$ is derived by $x_{adv}^{i}$ as the procedure illustrated in Fig.~\ref{fig:method}.

\subsection{Rationale behind the detection framework}
As the vocoder is data-driven and trained with genuine data during training, it models the distribution of genuine data, resulting in less distortion when generating genuine waveforms.
Thus, during inference, the vocoder's preprocessing will not affect the ASV scores of genuine samples too much, as reflected by the EER in the second row and last column of Table~\ref{tab:EER}.
However, suppose the inputs are adversarial samples. In that case, the vocoder will try to pull it back towards the manifold of their genuine counterparts to some extent, resulting in purifying the adversarial noise.

Take a non-target trial as an example, in which an ASV system should give the genuine sample a score below the threshold.
And after the nearly lossless reconstruction procedure (i.e., the yellow block and gray block in Fig.~\ref{fig:method}), the genuine sample will not change much, and the ASV score will remain largely unchanged.
In contrast to the genuine sample, the ASV score for the adversarial one is higher than the threshold.
And the reconstruction procedure will try to counter the adversarial noise, purify the adversarial sample, and decrease the ASV score for the adversarial sample. 
Then we can adopt the discrepancy of the score variations, $d_{adv}$ and $d_{gen}$, to discriminate between them, as shown in Fig.~\ref{fig:Hist-ep5}.
The transform, which makes $d_{gen}$ as small as possible while makes $d_{adv}$ as large as possible, is suitable for adversarial detection.

Also, the Griffin-Lim can be regarded as an imperfect transform as well, and it will also introduce distortion to affect the adversarial noise.
However, for genuine data, the distortion introduced by the Griffin-Lim is more significant than the vocoder, as it is not a data-driven method and can not be customized for a specific dataset, resulting in larger $d_{gen}$ and inferior detection performance.

%% file: 3-experimental-setup.tex
\section{Experimental setup}

\begin{table}[t]
\centering
\caption{EER with different $\epsilon$}
\scalebox{0.8}{
\begin{tabular}{cccccc}
\toprule
\multirow{2}{*}{Method} & \multicolumn{5}{c}{EER with different $\epsilon$ (\%)} \\
        & 20      & 15      & 10      & 5       & 0 (no attack)     \\ \hline
None    & 99.33   & 95.66   & 90.57   & 74.04   & 2.88     \\
Vocoder & 87.58    & 65.75   & 52.20   & 30.37   & 3.39    \\
GL-lin  & 95.23   & 80.83   & 66.73   & 39.49   & 3.93              \\
GL-mel  & 88.41   &65.39   & 49.76   & 26.67   & 3.81   \\ \bottomrule
\end{tabular}
}
\label{tab:EER}
\end{table}

\subsection{ASV setup}
The adopted system is a variation of X-vector system, and is modified from \cite{chung2020defence}.
We adopt the dev sets of Voxceleb1 \cite{nagrani2017voxceleb} and Voxceleb2 \cite{chung2018voxceleb2} for training.
Spectrograms are extracted with a Hamming window of width 25ms and step 10ms, and 64-dimensional fbanks are extracted as input features.
No further data augmentation and voice activity detection are adopted during training.
Cosine similarity is used for back-end scoring.
We adopt the trials provided in VoxCeleb1 test set for generating adversarial samples, evaluating the ASV performance and detection performance.

\subsection{Griffin-Lim and Parallel WaveGAN}
We use Griffin-Lim and Parallel WaveGAN in our experiments. The Griffin-Lim method, denoted as ``GL-lin", uses 100 iterations to reconstruct speech from linear spectrograms. ``GL-mel" denotes that linear spectrograms are first estimated from Mel-spectrograms using the pseudo inverse. Our Parallel WaveGAN method, denoted as "Vocoder", is modified from the public implementation\footnote[4]{https://github.com/kan-bayashi/ParallelWaveGAN}.
We use 80-dimension, band-limited (80-7600 kHz), and normalized log-mel spectrograms as conditional features. The window and hop sizes are set to 50 ms and 12.5 ms.
The architectures of the generator and discriminator follow those in \cite{yamamoto2020parallel}. We trained the model on the dev set of VoxCeleb1 ~\cite{nagrani2017voxceleb} for 1000k iterations, which takes around 5 days. 
Note that there is no overlap between the training data of Voxceleb1 for neural vocoder and the evaluation data of speaker verification.
To further show that the vocoder adopted in the proposed method is dataset independent, we also trained a universal vocoder~\cite{hsu2019towards} with the same structure as Vocoder, but on Lrg dataset \cite{hsu2019towards}, which is a large speech dataset containing 6 languages and more than 600 speakers.
The vocoder trained on Lrg is denoted as "Vocoder-L".

\subsection{ASV performance with genuine and adversarial inputs}
\label{subsec:ASV performance}
To evaluate the performance, we use the trials provide in VoxCeleb1 test set, which contains 37,720 enrollment-testing pairs.
During adversarial samples generation, $\alpha$ is set as $1$, attack budget $\epsilon$ is set as $5,10,15,20$. 
The adversarial attack is conducted in the time domain.
Also, note that it is time-consuming to generate adversarial samples.
We first evaluate the performance of our ASV system on genuine and adversarial samples.
The results are shown in the first row and the last column of Table~\ref{tab:EER}. "None" denotes that utterances are passed directly to the ASV system. 
We find that: 
(1) When testing on genuine samples, the ASV system achieved an EER of 2.88\%, comparable to recent ASV models.
When using generated speech as input, we found that the EER slightly increased.
(2) While introducing the adversarial attack, the EER increased from 2.88\% to over 70\%, which shows the effectiveness of the attack method.
The larger the attack budget $\epsilon$ is, the higher the attack intensity is.
One may question why the EER is over 50\%. The threshold of ASV is fixed, and the attackers try their best to do adversarial attack to make the score over and below the threshold for non-target and target trials respectively, resulting in the decisions for the trials reversed.

%% file: 4-experimental-results.tex
\section{Experimental Results}
\label{sec:expt-rst}

\begin{table}[t]
\centering
\caption{AUC with different $\epsilon$}
\label{tab:AUC}
\scalebox{0.8}{
\begin{tabular}{ccccc}
\toprule
\multirow{2}{*}{Method} & \multicolumn{4}{c}{AUC with different $\epsilon$ (\%)} \\
        & 20      & 15      & 10      & 5       \\ \hline
Vocoder & \textbf{99.94}    & \textbf{99.62}   & \textbf{99.12}   & \textbf{96.52}   \\
GL-lin  & 97.89   & 97.39   & 95.87   & 89.86   \\
GL-mel  & 99.01   & 97.64   & 95.41   & 87.52   \\ \bottomrule
\end{tabular}
\vspace{-5pt}
}
\end{table}

The defense performance is shown in Table~\ref{tab:EER}.
The EER decrease after audio re-synthesis illustrates that all the three methods can slightly alleviate the adversarial noise.
In contrast to \cite{joshi2021adversarial}, which effectively purifies the adversarial noise for speaker identification by vocoders, directly applying vocoders for defending speaker verification does not work.
Yet the re-synthesis process will not affect the genuine EER too much. 
Thus, we adopt the difference of the ASV scores between the original and re-synthesized audio as a good indicator to distinguish among genuine and adversarial samples. 

\begin{figure}[t]
\centering
\includegraphics[width=0.7\linewidth]{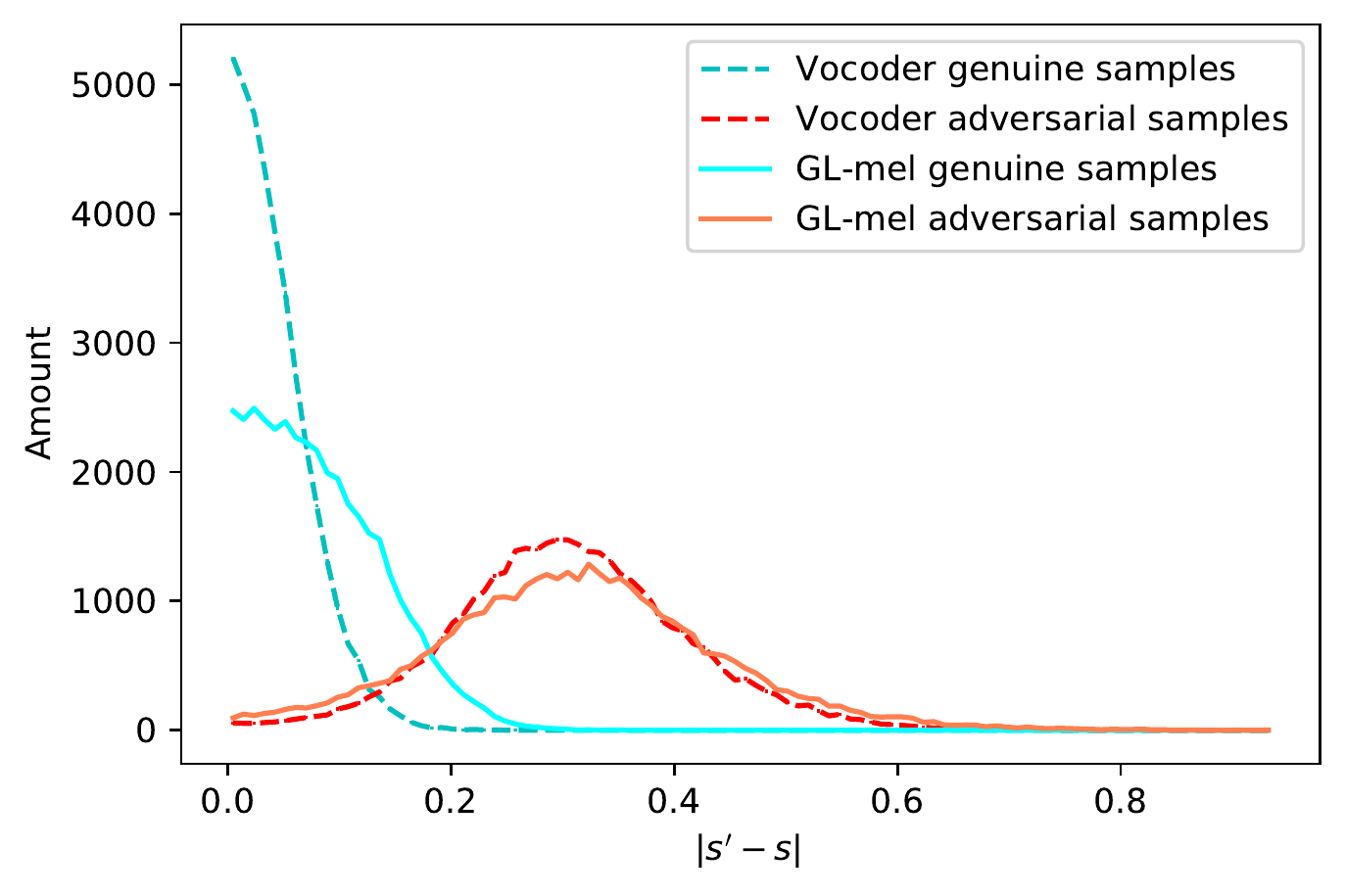}
\caption{$|s-s'|$ distribution with $\epsilon=10$}
\label{fig:Hist-ep5}
\end{figure}

As mentioned in Sec~\ref{subsec:vocoder based detection}, we use the discrepancy $|s-s'|$ to distinguish between genuine and adversarial samples.
The distributions of the discrepancy are in Fig.~\ref{fig:Hist-ep5}.
A threshold $\tau_{det}$ can be determined using genuine samples by Eq.~\ref{eq:det-threshold} to separate genuine and adversarial samples.
The detection framework based on vocoder in Fig.~\ref{fig:method} is effective, where the discrepancy $|s-s'|$ for genuine samples is small, yet $|s-s'|$ for adversarial samples is large.
Notice that the distribution overlap between the GL-mel genuine samples and adversarial samples is more considerable than that of Vocoder, as shown in Fig.~\ref{fig:Hist-ep5}.
As Vocoder is trained from genuine data, it attains the capacity of making $|s-s'|$ for genuine samples small enough.
Yet Griffin-Lim can only enlarge $|s-s'|$ for adversarial samples.

\begin{figure}[h]
    \centering
    \begin{minipage}[h]{0.23\textwidth}
        \centering
        \includegraphics[width=1.6in]{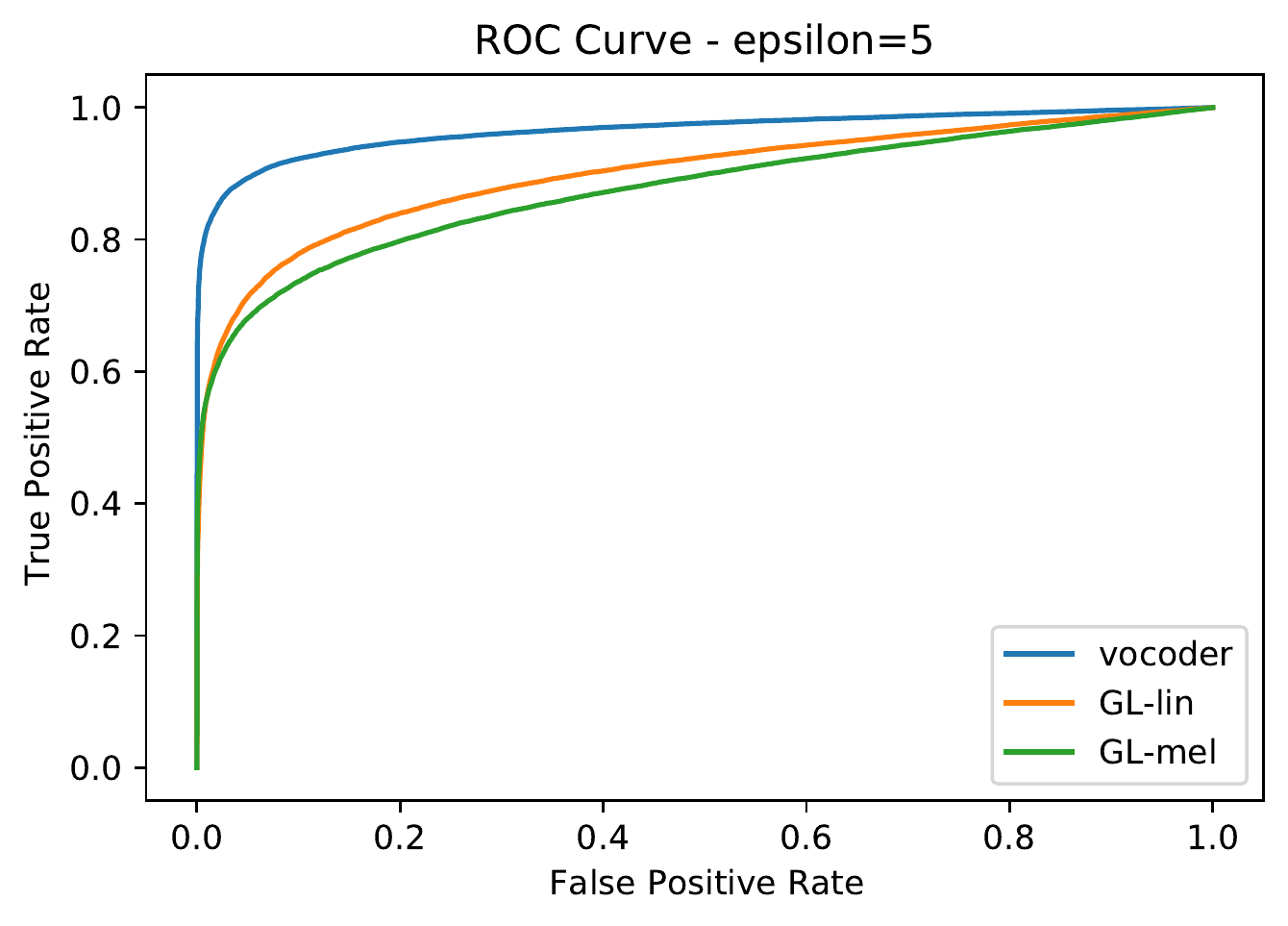}
        {\par \footnotesize (a)\par}
    \end{minipage}
    \hfill
    \begin{minipage}[h]{0.23\textwidth}
        \centering
        \includegraphics[width=1.6in]{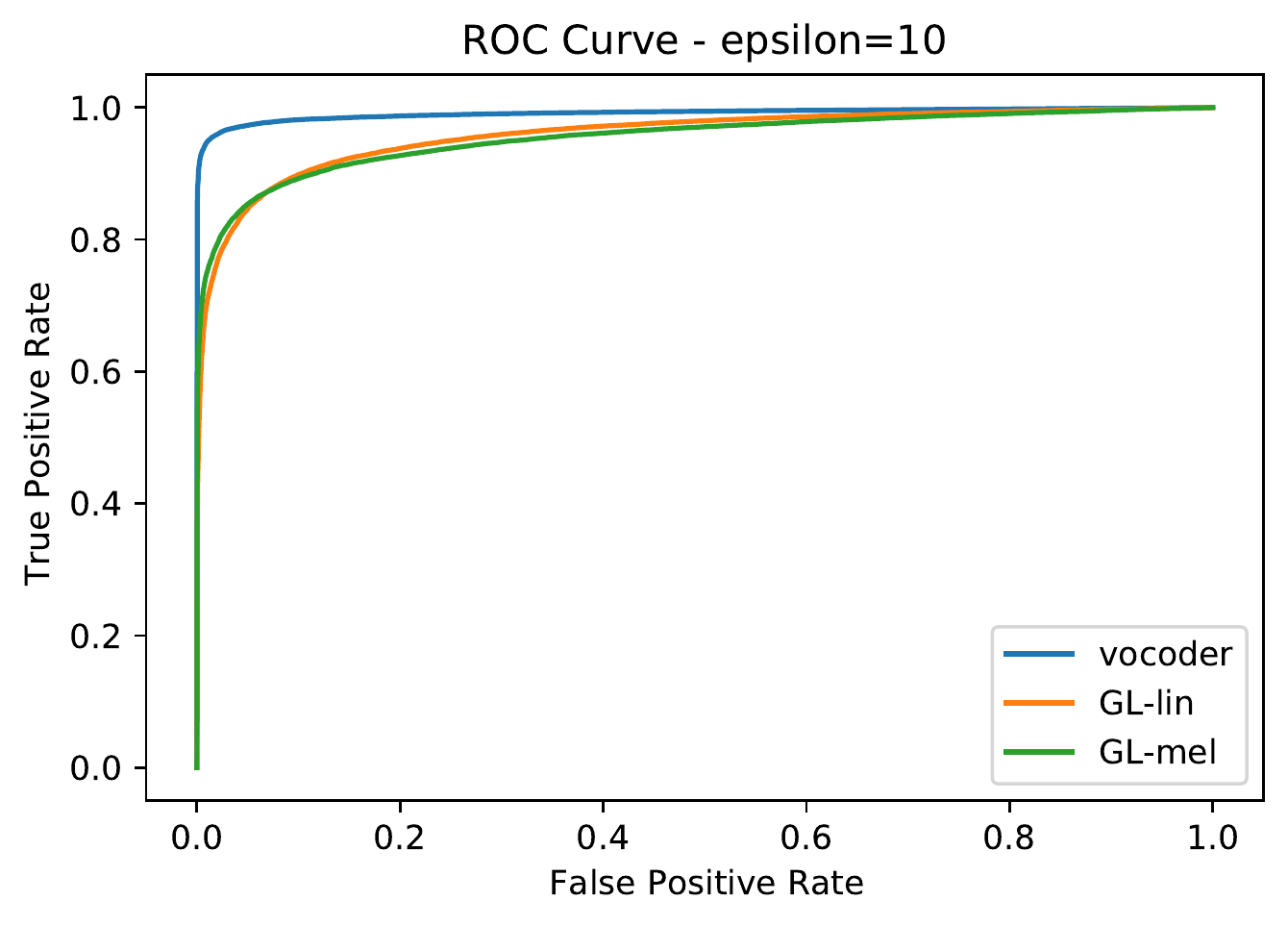}
        {\par \footnotesize (b)\par}
    \end{minipage}
    \hfill
    \begin{minipage}[h]{0.23\textwidth}
        \centering
        \includegraphics[width=1.6in]{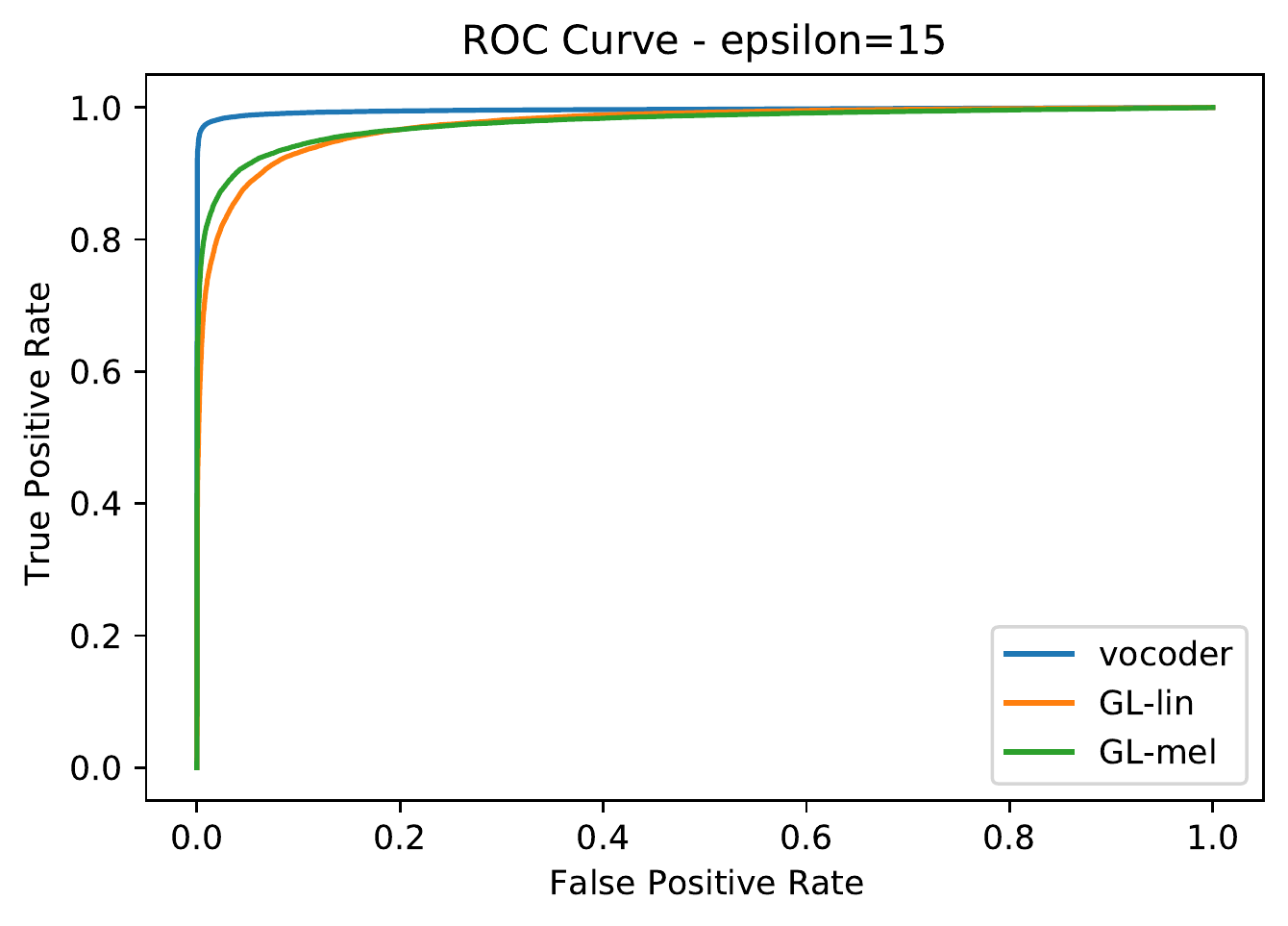}
        {\par \footnotesize (c)\par}
    \end{minipage}
    \hfill
    \begin{minipage}[h]{0.23\textwidth}
        \centering
        \includegraphics[width=1.6in]{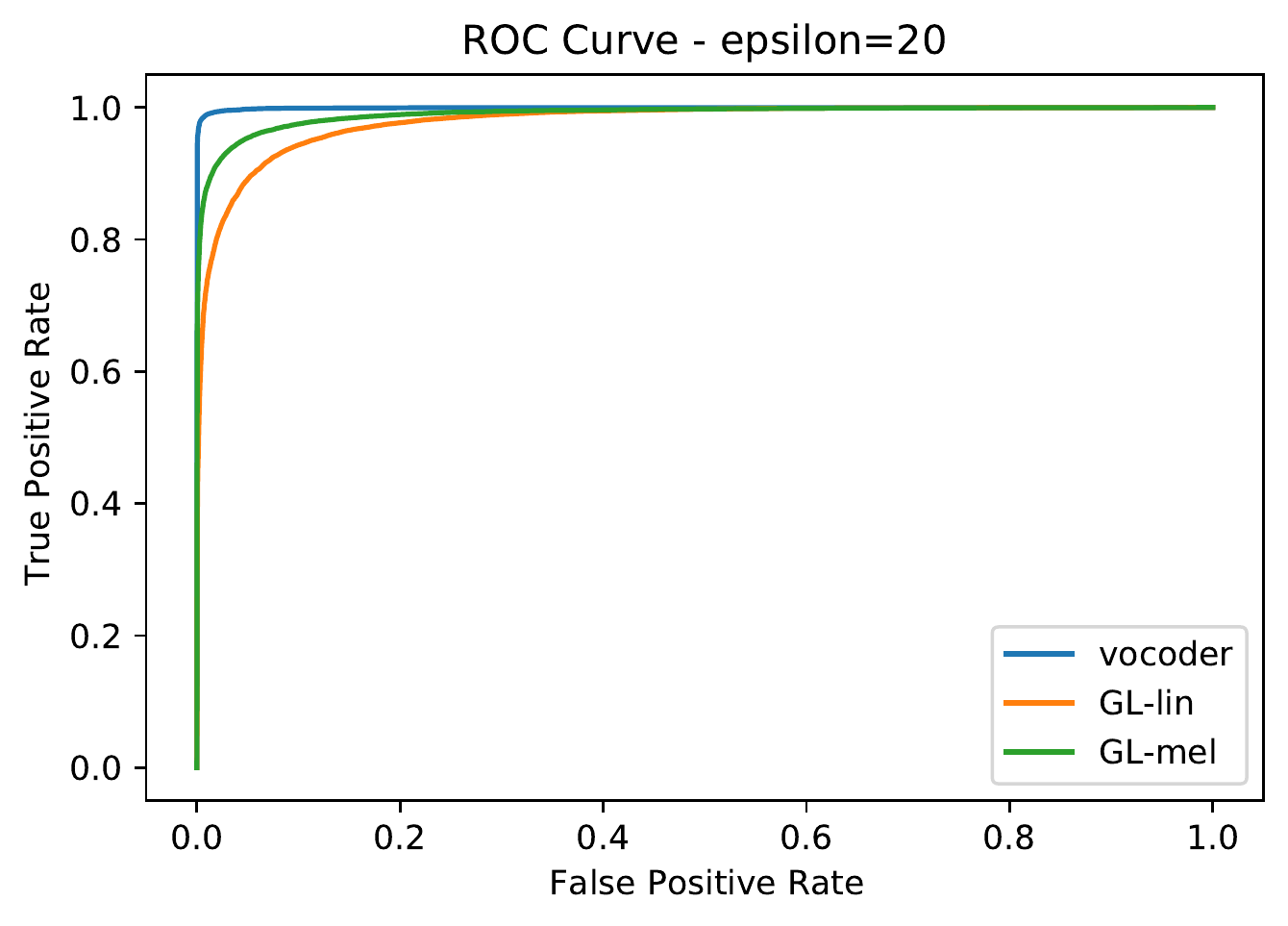}
        {\par \footnotesize (d)\par}
    \end{minipage}
    \vspace{-5pt}
    \caption{ROC curve under different epsilon ($\epsilon$) }
    \label{fig:roc-curve-epsilon=5}
    \vspace{5pt}
\end{figure}

Fig.~\ref{fig:roc-curve-epsilon=5} shows the receiver operating characteristic (ROC) curves of different methods.
The curves show the true positive rates for adversarial samples and false positive rates for genuine samples using different thresholds for detection.
The larger the area under the curve (AUC) is, the better the detection performance.
Both GL-lin and GL-mel achieve effective detection performance, and Vocoder performs better than them.
Table~\ref{tab:AUC} shows the AUC of different methods. 
Table~\ref{tab:AUC} indicates that the proposed method is powerful for adversarial samples detection as all AUCs are approaching or greater than 90\%.
Also, Vocoder outperforms GL-lin and GL-mel in all the settings.

\begin{table}[t!]
\centering
\caption{Detection rate with different $\epsilon$}
\label{tab:detection_rate}
\scalebox{0.7}{
\begin{tabular}{cccccc}
\toprule
\multirow{2}{*}{$FPR_{given}$} & \multirow{2}{*}{Method} & \multicolumn{4}{c}{Detection rate with different $\epsilon$ (\%)} \\
                        &           & 20        & 15        & 10        & 5         \\ \hline
                        & Vocoder   & \textbf{99.76}    & \textbf{98.82}    & \textbf{97.30}    & \textbf{89.33} \\
\multirow{3}{*}{0.05}   & Vocoder-L    & 99.38     & 97.23     & 94.07     & 81.21     \\
                        & GL-lin    & 89.12     & 88.30     & 84.64     & 71.29     \\
                        & GL-mel    & 95.39     & 91.33     & 85.37     & 68.07     \\ 
                        & Gaussian    & 34.54     & 51.29     & 61.56     & 68.57     \\ \hline
\multirow{3}{*}{0.01}   & Vocoder   & \textbf{98.92}    & \textbf{97.56}    & \textbf{94.76}    & \textbf{81.60} \\
                        & Vocoder-L    & 97.96     & 94.37     & 88.77     & 70.15     \\
                        & GL-lin    & 73.62     & 73.63     & 70.62     & 56.37     \\
                        & GL-mel    & 87.98     & 82.27     & 75.04     & 56.07     
                        \\ \hline
\multirow{3}{*}{0.005}  & Vocoder   & \textbf{98.30}    & \textbf{96.78}    & \textbf{93.25}    & \textbf{78.21} \\
                        & Vocoder-L    & 96.78     & 92.58     & 85.81     & 64.65     \\
                        & GL-lin    & 64.76     & 64.97     & 62.85     & 49.32     \\
                        & GL-mel    & 83.94     & 77.71     & 70.47     & 51.42     
                        \\ \hline
\multirow{3}{*}{0.001}  & Vocoder   & \textbf{96.04}    & \textbf{93.89}    & \textbf{88.60}    & \textbf{68.58} \\
                        & Vocoder-L    & 93.36     & 87.34    & 78.24     & 53.18     \\
                        & GL-lin    & 45.10     & 45.27     & 44.72     & 34.28     \\
                        & GL-mel    & 72.53     & 65.98     & 59.66     & 40.98     
                        \\ \bottomrule
\end{tabular}
}
\vspace{-5pt}
\end{table}

Table~\ref{tab:detection_rate} shows the detection results on adversarial samples with different $\epsilon$.
$FPR_{given}$ column lists different false acceptable rates.
The threshold $\tau_{det}$ was determined according to $FPR_{given}$ as shown in Eq.~\ref{eq:det-threshold}.
Gaussian denotes that we use Gaussian filter to replace feature extraction and vocoder (yellow block and gray block in Fig.~\ref{fig:method}).
Gaussian filter \cite{wu2020defense,wu2021adversarialasv} is usually adopted as an attack-agnostic method to counter adversarial samples, so we also set it as our baseline.
The observations and analysis are concluded as follows:
(1) We find that using Vocoder performs the best among all methods.
In most cases, more than 90\% of the adversarial samples could be detected. While with a large $\epsilon$ or $FPR_{given}$, all the detection rates even exceeded 95\%.
Even with a small $\epsilon$ of 5, the detection rates can still approach or exceed 80\%.
The results indicate that the proposed method can effectively detect adversarial samples.
(2) Gaussian based detection performs the worst, and even with $FPR_{given}=0.05$, the detection rates are still lower than vocoder based detection with $FPR_{given}=0.001$.
As it is not a comparable baseline, we do not show its results in other settings due to space limitation.
(3) For Griffin-Lim based methods, we find that they might be good approaches for detection with a large $\epsilon$ or $FPR_{given}$.
However, in stricter cases (smaller $\epsilon$ or $FPR_{given}$), the detection rates of GL-lin and GL-mel decrease drastically.
We argue that Griffin-Lim is a pseudo, nearly lossless transform, so we can, to some extent, adopt it to replace the vocoders in the adversarial detection framework in Fig.~\ref{fig:method}. 
While Griffin-Lim is not a data-driven method and can not model the genuine data manifold well, it results in higher $|s'-s|$ for genuine samples as shown in Fig.~\ref{fig:Hist-ep5}, and thus the detection performance is not comparable to Parallel WaveGAN.
(4) As shown in Table~\ref{tab:detection_rate}, the detection rate for the Vocoder-L is very close to Vocoder, which indicates the vocoder adopted for the proposed detection method is kind of dataset independent.

Also we try to use the vocoder to detect the Gaussian noise. 
Under the $FPR_{given}$ of 1\%, the detection rate for Gaussian noise is 0.97\%.
Due to space limitation, we won't show the details here.

%% file: 5-conclusion.tex
\section{conclusion}
\label{sec:conclusion}
This work adopts the neural vocoder to detect adversarial samples for ASV.
The proposed method accomplishes effective detection performance and outperforms the Griffin-Lim baseline in all the settings.
The key properties of the vocoder, including less distortion in genuine samples and better purification ability for adversarial samples, are both important for detection.
We will compare the performance of detecting adversarial samples with other attacking methods for ASV in our future work.
Also, we will evaluate the detection performance when the detection method is known to the attackers.